\documentclass[a4paper,reqno]{amsart}
\usepackage[margin=2.5cm,papersize={19.75cm,28cm}]{geometry}
\usepackage[all]{xy}           
\usepackage{amssymb}           
\usepackage{hyperref}
\usepackage{eucal}
\usepackage{epsfig}
\usepackage{graphicx}

\numberwithin{equation}{section}

\newtheorem{definition}{Definition}[section]

\newtheorem{theorem}[definition]{Theorem}

\newtheorem{remarkth}[definition]{Remark}

\renewcommand{\emph}[1]{{\bfseries\itshape{#1}}}

\newcommand{\R}{\mathbb{R}}      

\makeatother

\newcommand{\Su}{\mathbb{S}}

\begin{document}

\title{Quasivelocities and Optimal Control for underactuated Mechanical
 Systems
}

\author[L. Colombo]{Leonardo Colombo}
\address{L. Colombo:
Instituto de Ciencias Matem\'aticas  (CSIC-UAM-UC3M-UCM), Serrano
123, 28006 Madrid, Spain and Departamento de Matem\'atica,
Universidad Nacional de La Plata, Calle 50 y 115, La Plata, Buenos
Aires, Argentina} \email{leonardocolombo867@gmail.com}

\author[D.\ Mart\'{\i}n de Diego]{David Mart\'{\i}n de Diego}
\address{D.\ Mart\'{\i}n de Diego: Instituto de Ciencias Matem\'aticas (CSIC-UAM-UC3M-UCM), Serrano 123, 28006
Madrid, Spain} \email{david.martin@icmat.es}

\begin{abstract}
This paper is concerned with the application of the theory of quasivelocities for optimal
control  for underactuated mechanical systems. Using this theory, we convert the original problem in a variational second-order lagrangian system subjected to constraints.  The equations
of motion are geometrically derived using an adaptation of the classical Skinner and Rusk
formalism.
\end{abstract}

\maketitle


\section{Introduction}

 The mathematical activity in the last century in dynamical systems, mechanics and related areas has been recently extended to the control and optimal control of mechanical systems. In particular, there are an
increasing interest in the control of underactuated mechanical
systems (see \cite{bullolewis,cortes} and references therein). These type of
mechanical systems are  characterized by the fact that there are
more degrees of freedom than actuators, being their qualitative behavior quite different than
fully actuated control systems.

 Geometrically, quasivelocities are the components of velocities, describing a mechanical system,  relative to a set of vector fields (in principle, local)   that span on each point the fibers
of the tangent bundle of the configuration space. The main point is that these vector fields need not be
associated with (local) configuration coordinates on the configuration space.
In this paper we will use  quasivelocities as a tool for describing  optimal
control problem for underactuated mechanical systems.

The paper is organized as follows. In Section 2 we describe  the
concept of quasivelocities from a geometric point of view and
we introduce the Euler-Lagrange equations on quasivelocities (called Hamel equations). In section 3 we describe the conditions of optimality for  optimal control
problem using the Skinner and
Rusk formulation \cite{SR,Maria}.

\section{Quasivelocities}
Let $Q$ be a $n$ dimensional differentiable manifold, and $L:
TQ\rightarrow \mathbb{R}$ a  Lagrangian function determining the dynamics.
Let $(q^A)$, $1\leq A\leq n$, be local coordinates on $Q$ and choose a local basis of vector fields ${X_B}$ with $1\leq B\leq n,$ defined in the same coordinate neighborhood. The components of $X_B$ relative to the standard basis ${\frac{\partial}{\partial q^j}}$ will be denoted $X_B^A$, that is $X_B=X_B^A(q)\frac{\partial}{\partial  q^A}$.

 Let $(y^1,...,y^n)$ (the \textbf{quasivelocities}) be the components of a velocity vector $v$ on $TQ$ relative to the basis $X_B$, then $$v= y^BX_B(q)=y^BX_B^A(q)\frac{\partial}{\partial q^A}.$$
 Therefore, $\dot{q}^A= y^BX_B^A(q)$, then $$ L(q,\dot q) = L(q, y^BX_B^A(q)):=l(q,y).$$ On $TQ$ we have induced coordinates $\{(q^A, y^A)\mid A=1,...,n\}.$

 The Lie bracket of the vector fields $X_A$ is  $[X_A, X_B] = \mathcal{C}_{AB}^{D}X_D$,  where $\mathcal{C}_{AB}^{D}$ are called \textit{Hamel's transpositional symbols or structure coefficients}.

 Given a Lagrangian function $L:TQ\rightarrow\mathbb{R}$, the \textit{Euler-Lagrange equations in quasivelocities or  Hamel equations} are \begin{eqnarray*}\label{eq1}
\dot q^{A} &=& y^{B}X_{B}^{A}(q)\\
\frac{d}{dt}\left(\frac{\partial l}{\partial y^{A}}\right) &=&
\frac{\partial l}{\partial q^{B}}X_{B}^{A} -
\mathcal{C}_{AB}^{D}y^{B}\frac{\partial l}{\partial y^{D}}
\end{eqnarray*}
These equations were introduced by  \cite{Ham}  (see also \cite{NF}). It is interesting to note that these equations admit a nice, useful   and intrinsic interpretation in terms of mechanics on Lie algebroids (see \cite{LeMaMa}.

\section{Optimal Control for Underactuated Mechanical Systems}

We recall that a Lagrangian Control System is \textit{underactuated}
if the numbers of the control inputs is less than the dimension of
the configuration space. We assume, in the
sequel, that the  system is controllable \cite{Blo}.

Consider a lagrangian function $L:TQ  \rightarrow
\mathbb{R}$. Adding external forces and controlled forces
we have that the equations of motion are:
$$\frac{d}{dt}\left(\frac{\partial L}{\partial\dot{q}^{A}}\right) -
\frac{\partial L}{\partial q^{A}}=F_{A}+u_a\overline{X}_{A}^{a}$$
where $F=F^{A}(q,\dot{q})dq^{A}$ represents given external forces
and $\overline{X}^{a}=\overline{X}^{a}_{A}(q)dq^{a},$ $1\leq a\leq
m<n,$ the control forces.

Complete with independent 1-forms $\overline{X}^{\alpha}$ to obtain a local basis
$\{\overline{X}^{a}, \overline{X}^{\alpha}\}$ of $\Lambda^{1}Q$ and
take its dual basis that we denote by $\{X_{a}, X_{\alpha}\}$. Now, considering the quasivelocities induced by the
 local
basis $\{X_{a}, X_{\alpha}\}$, the \emph{control equations} are written as
\begin{eqnarray*}\label{eq2}
\dot q^{A} &=& y^{B}X_{B}^{A}(q)\\
\frac{d}{dt}\left(\frac{\partial l}{\partial y^{a}}\right) -
\frac{\partial l}{\partial q^{B}}X_{a}^{B} +
\mathcal{C}_{aB}^{D}y^{B}\frac{\partial l}{\partial
y^{D}} &=& F_{A}X_{a}^{A}+u_{a},\\
\frac{d}{dt}\left(\frac{\partial l}{\partial y^{\alpha}}\right) -
\frac{\partial l}{\partial q^{B}}X_{\alpha}^{B} +
\mathcal{C}_{\alpha B}^{D}y^{B}\frac{\partial l}{\partial y^{D}} &=&
F_{A}X_{\alpha}^{A}.
\end{eqnarray*}
where $1\leq a\leq m,$  $m+1\leq \alpha\leq n,$  and $u(t)=(u_1(t),...,u_m(t))\in U$
where  $U$ is and open subset of $\mathbb{R}^{m}$ containing $\bf{0}$.

For solving an optimal control problem we need to find a
trajectory $(q^A(t), u^a(t))$ (called an optimal curve) of the configuration variables and
control inputs satisfying the control equations
from given initial and final conditions: $(q^{A}(t_0),
 {y}^{A}(t_0)),$
$(q^{A}(t_f),  {y}^{A}(t_f))$  and minimizing the cost functional $$\mathcal{A}=\int_{t_0}^{t_f} C(q^{A}(t),
y^{A}(t), u^{a}(t))dt.$$

For other hand, a second order variational Lagrangian problem with constraints is given by $$\min_{q(\cdot)}\int_{0}^T L(q^{A},\dot{q}^A, \ddot{q}^A)\;dt$$ subject to the constraints $$\Phi(q^{A},\dot{q}^A, \ddot{q}^A)=0.$$ In the sequel we will show the equivalence of both theories (optimal control for underactuated systems and second order variational problems with constraints) under some regularity conditions (see \cite{Blo} and references therein).
Indeed, our initial
 optimal control problem is equivalent to the following
constrained variational problem
$$\hbox{ Minimize } \mathcal{\overline{A}}=\int_{t_0}^{t_f}\widetilde{L}\left(q^{A}(t), y^{A}(t), \dot{y}^{A}(t)\right)dt$$
subject to constraints $$\Phi^{\alpha}(q^{A}, y^{A}, \dot y^{A})
= \frac{d}{dt}\left(\frac{\partial l}{\partial y^{\alpha}}\right) -
\frac{\partial l}{\partial q^{B}}X_{\alpha}^{B} +
\mathcal{C}_{\alpha B}^{D}y^{B}\frac{\partial l}{\partial y^{D}} -
F_{A}X_{\alpha}^{A}=0,$$
and where $\widetilde{L}$ is defined as $$\widetilde{L}(q^{A}, y^{A},
\dot y^{A}) = C\left( q^A, y^A, \frac{d}{dt}\left(\frac{\partial l}{\partial
y^{a}}\right) - \frac{\partial l}{\partial q^{B}}X_{a}^{B}
+ \mathcal{C}_{a B}^{D}y^{B}\frac{\partial l}{\partial y^{D}} -
F_{A}X_{a}^{A}\right).$$

More, geometrically, we have that $(q^{A}, y^{A}, \dot{y}^{A})$ are
coordinates on $T^{(2)}Q$ (the second order tangent bundle)  and the constraints $\Phi^{\alpha}$
determine a submanifold $\mathcal{M}$ of $T^{(2)}Q$ and
 $\widetilde{L}: T^{(2)}Q\rightarrow\mathbb{R}$.

The canonical immersion $j_2: T^{(2)}Q\rightarrow T(TQ)$ in the
induced coordinates $(q^{A}, y^{A}, \dot{y}^{A})$ is
\[
\begin{array}{rcl}
 T^{(2)}Q&\stackrel{j_2}{\rightarrow}& TTQ\\
      (q^{A}, y^{A}, \dot y^{A})& \mapsto&(q^{A}, y^{A}, X^{A}_{B}y^{B},\dot y^{A})
\end{array}
 \]

Assume that the matrix $\left(\frac{\partial^{2}l}{\partial
y^{\alpha}\partial y^{\beta}}\right)_{m+1\leq\alpha,\beta\leq n}$ is
regular, then we can rewrite the constraints in the form
$\dot{y}^{\beta} = G^{\alpha}(q^{A}, y^{A}, \dot{y}^{a})$ and select coordinates $(q^{A}, y^{A}, \dot{y}^{a})$ on $\mathcal{M}$.

Hence,$(j_2)_{|{\mathcal M}}: {\mathcal M}\rightarrow T(T Q)$ is
\[
\begin{array}{rcl}
 {\mathcal M}&\stackrel{(j_2)_{|{\mathcal M}}}{\rightarrow}& TTQ\\
      (q^{A}, y^{A}; \dot y^{a} )& \mapsto&(q^{A}, y^{A}; X^{A}_{B}y^{B}, \dot{y}^a, G^{\alpha}(q^{A}, y^{A}, \dot y^{a}))
\end{array}
 \]

Let us define $\widetilde{L}_{\mathcal{M}}$ by
$\widetilde{L}_{\mathcal{M}} =
\widetilde{L}\mid_{\mathcal{M}}:\mathcal{M}\rightarrow\mathbb{R}$
and consider $W_{0} = \mathcal{M}\times_{TQ} T^{*}TQ$ with induced
coordinates $(q^{A}, y^{A}, \dot{y}^{a}, p_{A}, \widetilde{p}_{A}).$

Now, we will describe geometrically the problem based on the Skinner and
Rusk formalism (see \cite{SR}).

\begin{figure}[h]\label{figura1}
$$\xymatrix{
  &&W_0={\mathcal M} \times_{TQ} T^{*}(TQ) \ar[lld]_{pr_1} \ar[dd]^{\pi_{W_0,TQ}} \ar[rrd]^{pr_2}&&\\
  {\mathcal M} \ar[rrd]^{(\tau_{TQ})|_{{\mathcal M}}} && && T^*TQ \ar[lld]^{\pi_{T^*Q}} \\
  && TQ &&
   }$$
   \caption{Skinner and Rusk Formalism}
\end{figure}
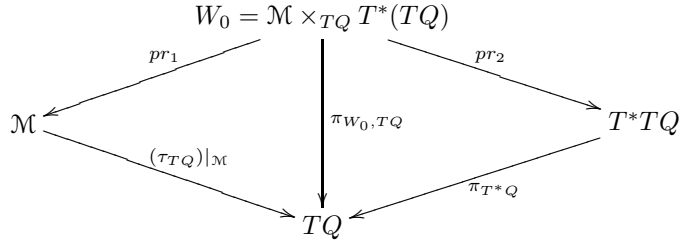

Let us define the 2-form $\Omega = pr_{2}^{*}(\omega_{TQ})$ on
$W_0$, where $\omega_{TQ}$ is the canonical symplectic form on
$T^{*}TQ$, and $\widetilde{H}(v_{x}, \alpha_{x})=\langle \alpha_{x},
(j_2)\mid_{\mathcal{M}}(v_{x})\rangle-\widetilde{L}_{\mathcal{M}}(v_x)$
where $x\in TQ, v_x\in
\mathcal{M}_{x}\cap (\tau_{TQ}\mid_{\mathcal{M}})^{-1}(x)$ and
$\alpha_{x}\in T^{*}_{x}TQ.$

In coordinates $$\Omega =dq^{A}\wedge dp_{A} + dy^{A}\wedge
d\widetilde{p}_{A},$$ $$\widetilde{H} = p_{A}X_{B}^{A}(q)y^{B} +
\widetilde{p}_{a}\dot{y}^{a} +
\widetilde{p}_{\alpha}G^{\alpha}(q^{A}, y^{A},
\dot{y}^{a})-\widetilde{L}_{\mathcal{M}}(q^{A}, y^{A},
\dot{y}^{a}).$$

The intrinsic expression of this constrained problem is given by the following presymplectic equation
\begin{equation}\label{leo}
i_{X}\Omega = d\widetilde{H}.
\end{equation}
Observe that  $ker\Omega = \displaystyle{\hbox{span }\langle\frac{\partial}{\partial\dot{y}^{a}}\rangle.}$

Following the Gotay-Nester-Hinds algorithm \cite{Gotay} for presymplectic hamiltonian systems we obtain the primary
constraints
$d\widetilde{H}\left(\frac{\partial}{\partial\dot{y}^{a}}\right) =
0,$ that is $$\varphi_{a} =
\frac{\partial\widetilde{H}}{\partial\dot{y}^{a}} =
\widetilde{p}_{a} + \widetilde{p}_{\alpha}\frac{\partial
G^{\alpha}}{\partial\dot{y}^{a}} -
\frac{\partial\widetilde{L}_{\mathcal{M}}}{\partial\dot{y}^{a}} =
0\; .$$

Therefore the dynamics is restricted to the manifold $W_1$
determined by the vanishing of the constraints $\varphi_{a}=0$.
Observe that $\dim W_1 = 4n$ with induced coordinates $(q^{A},
y^{A}, \dot{y}^{a}, p_{A}, \widetilde{p}_{\alpha}).$

A curve $t\rightarrow (q^{A}(t), y^{A}(t), \dot{y}^{a}(t), p_{A}(t),
\widetilde{p}_{A}(t))$ solution of the equations (\ref{leo}) must
verify the following system of differential-algebraic  equations.

\begin{eqnarray}
 \frac{dq^{A}}{dt}&=&X_{B}^{A}(q(t))y^{B}(t)\\
 \frac{dy^{\alpha}}{dt}&=&G^{\alpha}(q^{A}(t), y^{A}(t), \frac{dy^{a}}{dt}(t))\, ,\qquad
 \frac{dy^{a}}{dt} = \dot{y}^{a}(t)\label{qe-1}\\
 \frac{dp_A}{dt}&=&-p_{C}(t)\frac{\partial X_{B}^{C}}{\partial q^A}(q(t))y^{B}(t)-\widetilde{p}_{\alpha}(t)\frac{\partial G^{\alpha}}{\partial q^{A}}(q^{B}(t), y^{B}(t),\dot{y}^{b})\nonumber \\ && + \frac{\partial \widetilde{L}_\mathcal{M}}{\partial q^A}(q^{B}(t), y^{B}(t),\dot{y}^{b})\label{qe-2}\\
 \frac{d\widetilde{p}_{A}}{dt}&=&-p_C(t)X_{A}^{C}(q(t))-\widetilde{p}_{\alpha}(t)\frac{\partial G^{\alpha}}{\partial y^A}(q^{B}(t), y^{B}(t), \dot{y}^{b})\nonumber \\&&+ \frac{\partial \widetilde{L}_{\mathcal{M}}}{\partial y^A}(q^{B}(t), y^{B}(t), \dot{y}^{b})\label{qe-3}\\
\widetilde{p}_{a}(t) &=&-\widetilde{p}_{\alpha}(t)\frac{\partial
G^{\alpha}}{\partial\dot y^{a}}(q^{B}(t), y^{B}(t), \dot{y}^{b}) + \frac{\partial
\widetilde{L}_{\mathcal{M}}}{\partial \dot y^{a}}(q^{B}(t), y^{B}(t), \dot{y}^{b})\label{qe-4}
\end{eqnarray}

{}From Equations (\ref{qe-3}) and (\ref{qe-4}) we deduce

\begin{equation}\label{leoleoleo}
\frac{d}{dt}\left(\frac{\partial
\widetilde{L}_{\mathcal{M}}}{\partial \dot
y^{a}}-\widetilde{p}_{\alpha}\frac{\partial G^{\alpha}}{\partial\dot
y^{a}}\right) =-p_{C}X_{a}^{C}-\widetilde{p}_{\alpha}\frac{\partial
G^{\alpha}}{\partial y^a}+ \frac{\partial
\widetilde{L}_{\mathcal{M}}}{\partial y^a}
\end{equation}

Differentiating with respect to time, replacing in the previous
equality and using (\ref{qe-2}), we obtain the following system of equations
\begin{eqnarray}\label{eaea}
&&\frac{d^2}{dt^2}\left(\frac{\partial
\widetilde{L}_{\mathcal{M}}}{\partial \dot
y^{a}}-\widetilde{p}_{\alpha}\frac{\partial G^{\alpha}}{\partial\dot
y^{a}}\right) -\frac{d}{dt}\left(\frac{\partial
\widetilde{L}_{\mathcal{M}}}{\partial
y^a}-\widetilde{p}_{\alpha}\frac{\partial G^{\alpha}}{\partial
y^a}\right)\nonumber\\&&+
 X_{a}^{A}\left(\frac{\partial \widetilde{L}_{\mathcal{M}}}{\partial q^A}- \widetilde{p}_{\alpha}\frac{\partial G^{\alpha}}{\partial q^{A}}\right)-p_{C}y^{B}\left[X_{a}^{D}\frac{\partial X_{B}^{C}}{\partial q^D} - X_{B}^{D}\frac{\partial X_{a}^{C}}{\partial q^{D}}\right]=0
 \end{eqnarray}


Let us consider the 2-form $\Omega_{W_1}=i^{*}_{W_1}\Omega$ where
$i_{W_1}:W_1\hookrightarrow W_{0}$ is the canonical inclusion.

\begin{theorem}\label{teor} The submanifold $(W_1, \Omega_{W_1})$ is symplectic if and only if
for any system  of local coordinates  $(q^A,
y^A, \dot{y}^{a}, p_{A}, \widetilde{p}_{A})$ on $W_0$
$$\det({\mathcal R}_{ab})=\det\left(\frac{\partial^{2} \widetilde{L}_{\mathcal{M}}}{\partial
\dot
y^{a}\partial\dot{y}^{b}}-\widetilde{p}_{\alpha}\frac{\partial^{2}
G^{\alpha}}{\partial\dot
y^{a}\partial\dot{y}^{b}}\right)_{m\times m}\neq 0 \hbox{    along   } W_1.$$
\end{theorem}

\textbf{Proof:} Let us recall that $\Omega_{W_1}$ is symplectic of
and only if $T_{x}W_{1}\cap \left(T_{x}W_1\right)^{\perp} = 0 \ \
\forall x\in W_1$,
   where
   \[
   \left(T_{x}W_1\right)^{\perp}=\left\{ v \in T_xW_0 \ / \ \Omega_{W_0}(x)(v,w)=0, \hbox{for all } w\in T_x W_1\right\}.
   \]

   Suppose that $(W_1, \Omega_{W_1})$ is symplectic and that
   \[
   \lambda^a {\mathcal R}_{ab}(x)=0 \hbox{  for some  } \lambda^a\in \mathbb{R} \hbox{  and  } x\in W_1\; .
   \]
   Since
    \[
   \lambda^b {\mathcal R}_{ab}(x)=\lambda^b  d\varphi_a(x) \left(\left.\frac{\partial}{\partial \dot{y}^b}\right|_{x}\right)=0\; ,
   \]
   then, $\lambda^b \left.\frac{\partial}{\partial \dot{y}^b}\right|_{x}\in T_x W_1$ but it is also in $T_xW_1^\perp$. This implies that $\lambda_b=0$ for all $b$ and that the matrix $({\mathcal R}_{ab})$ is regular.

  Now, suppose that the matrix  $({\mathcal R}_{ab})$ is regular.
As we have observed
  \[
  d\varphi_a(x) \left(\left.\frac{\partial}{\partial \dot{y}^b}\right|_{x}\right)={\mathcal R}_{ab}(x)
  \]
  and therefore, $\left.\frac{\partial}{\partial \dot{y}^b}\right|_{x}\notin T_x W_1$ and, in consequence,
  \[
  T_xW_1\oplus \hbox{span } \left\{ \left.\frac{\partial}{\partial \dot{y}^b}\right|_{x}\right\}=T_x W_0.
  \]
  Now, let $Z\in T_{x}W_{1}\cap \left(T_{x}W_1\right)^{\perp}$ with $x\in W_1$. It follows that
  \[
  0=i_Z\Omega_{W_0}(x)\left(\left.\frac{\partial}{\partial \dot{y}^a}\right|_{x}\right), \hbox {  for all  } a   \hbox{ and  } i_Z\Omega_{W_0}(x)(\bar{Z})=0, \hbox{  for all  } \bar{Z}\in T_xW_1
  \]
   Then, $Z\in \ker \Omega_{W_0}(x)$. This implies that
   \[
   Z=\lambda_b \left.\frac{\partial}{\partial \dot{y}^b}\right|_{x}
   \]
   Since $Z\in T_xW_1$ then
   \[
   0=d \varphi_a(x)(Z)=d \varphi_a(x)\left(\lambda_b \left.\frac{\partial}{\partial \dot{y}^b}\right|_{x} \right)=\lambda_b{\mathcal R}_{ab}
   \]
    and, consequently, $\lambda_b=0$, for all $b$, and $Z=0$. \hfill $\Box$

Under the hypothesis of Theorem \ref{teor}, we can rewrite the necessary conditions for optimality as an explicit system of differential equations where Equation (\ref{eaea}) is replaced by
\begin{eqnarray}
\frac{d^3 y^a}{dt^3}&=& \Gamma^a\left( q^A,  y^A, \frac{d
y^a}{dt}, \frac{d^2 y^a}{dt^2}, p_A, \tilde{p}_{\alpha}\right)
\label{poi-3}
\end{eqnarray}

\subsection{Example}[The Planar Rigid Body]

The configuration space for this system in $Q=\R^2\times S^1$ and it can be considered as the simplest example in the category of rigid body dynamics.
The three degrees of freedom describe the  translations in $\R^2$ and the rotation about its center of mass. The configuration is given by the followings variables: $\theta$ describes the relative orientation  the body reference frame with respect to the inertial reference frame. The vector $(x,y)$ denotes the position of the center of mass measured with respect to the inertial reference frame.
The lagrangian is of kinetical type  $$L=\frac{1}{2}\dot{q}^{T}\mathcal{G}(q)\dot{q}, \hbox{ where } \mathcal{G}(q)=\left(
  \begin{array}{ccc}
  m & 0 & 0\\
   0& m & 0\\
  0& 0 & J\\
  \end{array}
 \right),$$
and where $m$ is the mass of the body and $J$ is its moment of inertia about the center of mass. If we assume that the body moves in a plane perpendicular to the direction of the gravitational forces being the potential energy  zero. For the planar body, the control forces that we consider are applied to a point on the body with distance ${h}>0$ from the center of mass, along the body $x-$axis (see \cite{bullolewis} for more details about this example).

The equations of motion are
\begin{eqnarray*}
m\ddot{x}&=&u_1\cos\theta-u_2\sin\theta\\
m\ddot{y}&=&u_1\sin\theta+u_2\cos\theta\\
J\ddot{\theta}&=&-h u_2
\end{eqnarray*}

The control fields are
\begin{eqnarray*}
X_1&=&\frac{\cos\theta}{m}\frac{\partial}{\partial x} + \frac{\sin\theta}{m}\frac{\partial}{\partial y}\\
X_2&=&-\frac{\sin\theta}{m}\frac{\partial}{\partial x}+\frac{\cos\theta}{m}\frac{\partial}{\partial y}-\frac{h}{J}\frac{\partial}{\partial\theta},
\end{eqnarray*}
and we complete the basis of vector fields  with $X_3= h\sin\theta\frac{\partial}{\partial x} - h\cos\theta\frac{\partial}{\partial y} -\frac{\partial}{\partial\theta}$

The nonzero structure functions are
\begin{eqnarray*}
\mathcal{C}_{12}^2&=&\frac{h}{mh^2+J}=-\mathcal{C}_{21}^2,\qquad
\mathcal{C}_{12}^3=-\frac{h^2}{(mh^2+J)J}=-\mathcal{C}_{21}^3\\
\mathcal{C}_{23}^1&=&-\frac{mh^2+J}{J}=-\mathcal{C}_{32}^1, \qquad
\mathcal{C}_{13}^{2}=\frac{J}{mh^2+J}=-\mathcal{C}_{31}^2\\
\mathcal{C}_{13}^3&=&-\frac{h}{mh^2+J}=-\mathcal{C}_{31}^3
\end{eqnarray*}

Taking the corresponding  quasivelocities $\{y^1, y^2, y^3\}$, we have that
\begin{eqnarray*}
\dot{x}&=&y^1\frac{\cos\theta}{m}-y^2\frac{\sin\theta}{m}+y^3 h\sin\theta\\
\dot{y}&=&y^1\frac{\sin\theta}{m}+y^2\frac{\cos\theta}{m}-y^3 h\cos\theta\\
\dot{\theta}&=&-y^2\frac{h}{J}-y^3\; .
\end{eqnarray*}

The Lagrangian of this system is $$l(x, y, \theta, y^1, y^2, y^3)=\frac{1}{2}\left[\frac{1}{m}(y^1)^2+ \frac{mh^2+J}{mJ} (y^2)^2 + (mh^2+J) (y^3)^2\right],$$  then the  Hamel equations with controls are:

\begin{eqnarray*}
u_1&=&\dot{y}^1+\frac{h}{J}(y^2)^2-hm(y^3)^2+ \frac{J-mh^2}{J}y^2y^3\\
u_2&=& \frac{J+mh^2}{J}\dot{y}^2--\frac{h}{J}y^1y^2-y^1y^3\\
0&=&(J+mh^2)\dot{y}^3+\frac{h^2}{J}y^1y^2+hy^1y^3\; .
\end{eqnarray*}

Consider the following cost functional
$$\mathcal{A} = \frac{1}{2}\int_{0}^{T}\left( u_{1}^{2} + u_{2}^{2}\right)\, dt.$$

Following our formalism this  optimal control problem is equivalent
to the constrained second-order variational problem determined by:
$$\widetilde{\mathcal{A}} = \int_{0}^{T} \widetilde{L}(x, y, \theta, y^1, y^2, y^3, \dot{y}^1, \dot{y}^2, \dot{y}^3)\, dt$$
and the second-order constraint
$$\Phi(x, y, \theta, y^1, y^2, y^3, \dot{y}^1, \dot{y}^2, \dot{y}^3) =(J+mh^2)\dot{y}^3+\frac{h^2}{J}y^1y^2+hy^1y^3= 0\; ,$$ where \begin{eqnarray*}
\widetilde{L}(x, y, \theta, y^1, y^2, y^3, \dot{y}^1, \dot{y}^2, \dot{y}^3) &=&\frac{1}{2}\left[\dot{y}^1+\frac{h}{J}(y^2)^2-hm(y^3)^2+ \frac{J-mh^2}{J}y^2y^3 \right]^2\\
&+&\frac{1}{2}\left[\frac{J+mh^2}{J}\dot{y}^2--\frac{h}{J}y^1y^2-y^1y^3\right]^2
\end{eqnarray*}

Now, we rewrite the second-order constraint in the form
\[\displaystyle{\dot{y}^3 =-\frac{h^2}{(J+mh^2)J}y^1y^2-\frac{h}{(J+mh^2)}y^1y^3\; .}
\]

Take now  $W_0 = {\mathcal M}\times T^{*}(T(\R^2\times \Su^{1}))$ with
coordinates $(x, y, \theta, y^1, y^2, y^3, \dot{y}^1 , \dot{y}^2,
p_{1}, p_{2}, p_{3}, \tilde{p}_{1}, \tilde{p}_{2}, \tilde{p}_{3})$.


Now, the presymplectic  $2$-form $\Omega$, the Hamiltonian
$\widetilde{H}$ and the primary constraints, $\varphi_{1}, \varphi_{2}$, are:
\begin{eqnarray*}
\Omega &=& dx\wedge dp_{1} + dy\wedge dp_{2}+ d\theta\wedge dp_{3} + d y^1\wedge d \tilde{p}_{1} + dy^2\wedge d\tilde{p}_{2}+ dy^3\wedge d\tilde{p}_{3}\; ,\\
\widetilde{H} &=&p_{1}\left[y^1\frac{\cos\theta}{m}-y^2\frac{\sin\theta}{m}+y^3 h\sin\theta\right]+ p_{2}\left[y^1\frac{\sin\theta}{m}+y^2\frac{\cos\theta}{m}-y^3 h\cos\theta\right]\\
&& -  p_{3}\left[y^2\frac{h}{J}+y^3\right] +
\tilde{p}_{1}\dot{y}^1 + \tilde{p}_{2}\dot{y}^2 - \tilde{p}_{3}\left(\frac{h^2}{(J+mh^2)J}y^1y^2+\frac{h}{(J+mh^2)}y^1y^3\right) \\
&&-
\frac{1}{2}\left[\dot{y}^1+\frac{h}{J}(y^2)^2-hm(y^3)^2+ \frac{J-mh^2}{J}y^2y^3 \right]^2-\frac{1}{2}\left[\frac{J+mh^2}{J}\dot{y}^2--\frac{h}{J}y^1y^2-y^1y^3\right]^2\\
\varphi_{1} &=& \frac{\partial\widetilde H}{\partial\dot{y}^1} =
\tilde{p}_{1} - \left[\dot{y}^1+\frac{h}{J}(y^2)^2-hm(y^3)^2+ \frac{J-mh^2}{J}y^2y^3 \right]=
0,\\
\varphi_{2} &=& \tilde{p}_{2} -\frac{J+mh^2}{J} \left[\frac{J+mh^2}{J}\dot{y}^2--\frac{h}{J}y^1y^2-y^1y^3\right]\; ,
 \end{eqnarray*}
 i.e.,
 \begin{eqnarray*}
 \tilde{p}_{1}&=&\dot{y}^1+\frac{h}{J}(y^2)^2-hm(y^3)^2+ \frac{J-mh^2}{J}y^2y^3\\
 \tilde{p}_2&=&\frac{J+mh^2}{J} \left[\frac{J+mh^2}{J}\dot{y}^2--\frac{h}{J}y^1y^2-y^1y^3\right]\; .
 \end{eqnarray*}

These constraints determine the submanifold $W_1$. Applying
the Theorem \ref{teor}, we deduce that the 2-form
 $\Omega_{W_1}$, restriction of $\Omega$ to $W_1$, is  symplectic since
\[
\left(
\begin{array}{rr}
{\mathcal R}_{11}&{\mathcal R}_{12}\\
{\mathcal R}_{21}&{\mathcal R}_{22}
\end{array}
\right)=
\left(
\begin{array}{rr}
1&0\\
0&(J+mh^2)/J
\end{array}
\right)
\]
is regular.
%


Therefore, the algorithm stabilizes in the first constraint
submanifold $W_1$. Moreover, there exists a unique solution of the
dynamics, a vector field  $X$ which satisfies $i_{X}\Omega_{W_{1}} =
d\widetilde{H}_{W_{1}}$.
 In consequence, we have a unique control input which extremizes the objective function ${\mathcal A}.$
If we take the flow $F_{t}: W_{1}\rightarrow W_{1}$ of the vector
field $X$ then we have that $F_{t}^{*}\Omega_{W_1} = \Omega_{W_1}$, then the evolution is symplectic preserving.
Obviously, the hamiltonian function $\widetilde{H}\big{|}_{W_1}$
is preserved by the solution of the optimal control problem, that is
$\widetilde{H}\big{|}_{W_1}\circ F_t=\widetilde{H}\big{|}_{W_1}$.
Both  properties, symplecticity and preservation of energy, are
important geometric invariants. In \cite{ldm}, we construct,
using discrete variational calculus, numerical integrators which
inherit some of the geometric properties of the optimal control
problem (symplecticity, momentum preservation and a very good energy
behavior).

\end{document}